\documentclass[12pt]{article}
\usepackage{graphicx}
\newlength{\vshift}
\newlength{\hshift}

\usepackage{amssymb}
\usepackage{amsmath}
\usepackage{latexsym}
\usepackage{amsthm}
\usepackage{amsopn}
\usepackage{xcolor}
\usepackage{cite}
\usepackage{float}
\usepackage{float}

\begin{document}
\vspace*{3cm}
\begin{center}
{\bf{\Large  Effects of minimal length on Berry phase and spin-orbit interactions}}
\vskip 4em{ {\bf S. Aghababaei $^{a,}$}\footnote{s.aghababaei@yu.ac.ir}\:, {\bf H. Moradpour $^{b,}$}\footnote{hn.moradpour@maragheh.ac.ir}\:, \: {\bf G. Rezaei $^{a,}$}\footnote{grezaei@yu.ac.ir}\: and \bf S. Khorshidian$^{a}$}
\vskip 1em
$^a$Department of Physics, Faculty of Sciences, Yasouj University, 75918-74934, Yasouj, Iran\\
$^b$Research Institute for Astronomy and Astrophysics of Maragha (RIAAM), University of Maragheh, P.O. Box 55136-553, Maragheh, Iran\\
\vspace*{1.6cm}

\begin{abstract}
The effect of Generalized Uncertainty Principle (GUP) on Berry phase is studied using the perturbation approach and up to the first order of approximation. Thereinafter, the obtained results are extended to a quantum ring in which two types of spin-orbit interactions, including Rashba and Dresselhaus interactions, can be felt by electrons. Comparing the final results with the accuracy of Berry phase detectors, one can find an upper bound on GUP parameter as $\beta_{0}<10^{46}$ and $\beta_{0}<10^{51}$ from Rashba and Dresselhaus interactions, respectively, in agreement with previous results.
\end{abstract}
\end{center}
\section{Introduction}
~~Attempt for reconciling gravity and quantum mechanics is one of the important subjects in theoretical physics which has not yet found a satisfactory solution. Several studies of quantum gravity theories such as string theory \cite{stringHUP} and loop quantum gravity \cite{loopHUP} propose to extend the standard Heisenberg Uncertainty Principle (HUP), which incorporates gravitational phenomena at higher energies and converges to Heisenberg’s uncertainty principle at lower energies, as follows:
\begin{eqnarray}
(\Delta X)(\Delta P)\geq\frac{\hbar}{2}(1+\beta (\Delta P)^{2}+...),
\label{HUP}
\end{eqnarray}
where $ \beta $ is a small deformation parameter with dimension of $(\text{Length})^{2}$ in the natural unit. It is also customary to define dimensionless parameter $\beta_0$ through writing $\beta=\dfrac{\beta_{0}}{(M_{P}c)^{2}}$ where $M_{P}$ denotes Planck mass. This Generalized Uncertainty Principle (GUP) includes a nonzero minimal uncertainty in position (minimal length), given by $(\Delta x)_{min}=\hbar\sqrt{\beta}$, and is of the order of Planck length ($10^{-35} m$) \cite{minimallength}. In recent years, the effects of GUP on various physical systems have been studied by many authors in both low energy \cite{lowGUP} and high energy \cite{highGUP} regimes. Current experiments can also set upper bounds on $\beta_0$. It can be constrained by using kinds of phenomenological approaches which can be summarized as follows: the best bounds from the non-gravitational origin are in the range of $\beta_{0}<10^{6}$ \cite{Bushev2019}; the best upper bound with gravitational origin (with violation of the equivalence principle) is $\beta_{0}<10^{21}$ \cite{Ghosh2014}; and for the best gravitational bound with respecting the equivalence principle and using gravitational wave event GW150914, we have $\beta_{0}<10^{60}$ \cite{Feng2017}. Motivated by string theory models, some authors also find some estimations the order of unit for the GUP parameter \cite{theo1, theo2}. There is a gap between theoretical predictions and experimental bounds which requires a big leap in the experimental techniques to probe a vast region for $\beta_{0}$ where the current upper bounds are in range of $10^{6}$ \cite{Bushev2019} to $10^{78}$\cite{78bound}. Indeed, in this regard, various quantum systems have been proposed to investigate the possibility of detecting the GUP effects which finally help us in getting some upper bounds on the GUP parameter \cite{lowGUP, highGUP, Bushev2019, Ghosh2014, Feng2017, theo1, theo2, 78bound, 1, 2, 5, GUPwork}.
\par
On the other hand, the geometrical phase proposed by Berry in adiabatic cyclic process, contains information on the geometrical properties of the parameter space of a quantum system \cite{berryphase}. Berry phase can play a fundamental role in understanding the behavior of a variety of systems and phenomena, and thus, it is interesting to develop experimental probes to measure Berry phase, as well as theoretical models that connect its behavior to microscopic information or external fields \cite{ABphase,BP1}. In recent years, the spin-dependent transport experiments have demonstrated that it is possible to control the geometric phase of electrons by the application of in-plane fields in semiconductor devices such the quantum rings built on In-Ga-As structures \cite{In-Ga-As}. In such devices, when an electron is transmitted from the source to drain, the spin precession and the spin-dependent phase (spin Berry phase) of the electron are controllable by the Rashba spin-orbit coupling, while this effect is regulated by a perpendicular electric field \cite{spinberry}. In addition, both the Aharonov-Bohm (AB) \cite{ABphase} and Aharonov-Casher (AC) \cite{ACphase} effects have been studied for quantum rings both experimentally and theoretically and may also be useful for controlling the spins of electrons \cite{spinAC}. The purpose of this work is to explore the geometrical Berry phase in the presence of GUP. In this regard, we show that the GUP effect provides the new contributions to Berry phase for electrons confined in a ring. We express that the geometrical phase produced by spin-orbit interactions in condensed matter systems can generate a way to follow GUP effects. It is also addressed that the current experiments in the solid material indicate the best upper bound on the GUP parameter.
\par
The paper is organized as follows:
Section II describes a generalized framework for the GUP effect. Section III is devoted to deriving Berry phase with the GUP correction. A discussion of GUP effects on the geometrical Berry factor for electrons inside a quantum ring is presented in section IV. The geometrical phase to the electron in a ring are discussed by the Rashba and Dresselhaus interactions. In section V, we summarize our results and conclusions.
\section{Generalized framework}
~~A general deformed Heisenberg algebra has been proposed in the variety of models of quantum gravity which are predicted a leading quadratic term in the momenta type correction to the HUP by Kempf, Mangano, and Mann \cite{HUP}. The existence of GUP leads to fundamental consequences on the mathematical basis of quantum mechanics. One of its most important implications is the deformation of commutation relation between position and momentum operators \cite{HUP}, as follows:
\begin{eqnarray}
[X,P]=i\hbar(1+\beta P^{2}).
\label{GUP}
\end{eqnarray}
~~Various topics such as harmonic oscillator \cite{1}, hydrogen atom \cite{2}, particle in a gravitational quantum well \cite{3}, have recently been studied using this modified version of quantum mechanics and its effects on corresponding Schr\"{o}dinger equations. In the relativistic regimes, it was used in studying the Dirac oscillator \cite{4}, the Klein-Gordon equation \cite{5}, the Casimir effect and the black body radiation \cite{6}. In general, GUP modifies Heisenberg algebra \cite{HUP, HUP2} as form of:
\begin{eqnarray}
&[X_{i},P_{j}]&=i\hbar\{(1+\beta P^{2})\delta_{ij}+\beta'P_{i}P_{j}\},\nonumber\\
&[P_{i},P_{j}]&=0,\nonumber\\
&[X_{i},X_{j}]&=i\hbar\dfrac{2\beta-\beta'+\beta(2\beta+\beta')P^{2}}{1+\beta P^{2}}(P_{i}X_{j}-X_{i}P_{j}),
\label{GA}
\end{eqnarray}
where $\beta$ and $\beta'$ are the GUP parameters, producing two versions of deformed quantum mechanics. First, the momentum representation is given in Ref. \cite{HUP2} by
\begin{eqnarray}
X_{i}&=&i\hbar\bigg((1+\beta p^{2})\dfrac{\partial}{\partial p_{i}}+\beta'p_{i}p_{j}\dfrac{\partial}{\partial p_{j}}+\gamma p_{i}\bigg),\nonumber\\
P_{i}&=&p_{i}.
\label{mumentumspace}
\end{eqnarray}
~~Here, operators $x_{i}$ and $p_{i}$ satisfy the standard commutation relations of ordinary quantum mechanics (they are canonical operators), and $\gamma$ is a parameter related to $\beta$ and $\beta'$. The solution of the deformed Schr\"{o}dinger equation in this approach is not often simple and there are only a few problems which have been solved exactly in the momentum approach with a minimal length in the case of $\beta'=2\beta$ \cite{2}. The other important representation is the following position representation \cite{7}
\begin{eqnarray}
X_{i}&=&x_{i},\nonumber\\
P_{i}&=&p_{i}(1+\beta p^{2})+\mathcal{O}(\beta^{2}),
\label{coordinatespace}
\end{eqnarray}
which is valid in the case $\beta'= 2\beta$ up to the first order of $\beta$. The simplicity of applying the perturbation theory to solve the Schr\"{o}dinger equation modified by GUP is the main advantage of the position representation \cite{7}. Hence, we use this representation in this work.
\section{Berry phase with a GUP}
~~In the early 1980s, it was shown that a quantum mechanical system acquires a geometric phase for a cyclic motion in the parameter space. This geometric phase under adiabatic motion is called Berry phase \cite{berryphase} while its generalized, which encompasses the non-adiabatic motion is known as the Aharonov-Anandan phase \cite{AAphase}. A manifestation of the Berry phase is the well-known Aharonov-Bohm (AB) phase \cite{ABphase} of an electrical charge which cycles around a magnetic flux. Aside from the AB effect, the first experimental observation of Berry phase was reported in 1986 for photons in a wound optical fiber \cite{expberry}. Another important
Berry phase effect is the Aharonov-Casher (AC) effect \cite{ACphase}, which has been proposed to occur when an electron with a spin-orbit interaction propagates in a ring structure under the effect of an external magnetic field perpendicular to the ring plane. In solid materials, especially for those with large spin-orbit coupling, the spin Berry phase has manifested by the various quantum phenomena, including anomalous Hall effect, spin Hall effect, valley Hall effect, anomalous thermoelectric effect, electronic polarization, orbital magnetization, magnetoresistance, magneto-optic effect, and 3D/2D topological insulator \cite{so}.
\par
To find the GUP effects on geometrical Berry phase, let us firstly consider usual quantum mechanics with general Hamiltonian
\begin{eqnarray}
 H(p,x)=\frac{p^{2}}{2m}+V(x),
 \label{H0}
\end{eqnarray}
where the corresponding Schr\"{o}dinger equation is given by
\begin{eqnarray}
H(p,x)|\phi_{n}>=E_{n}^{(0)}|\phi_{n}>.
\label{sheq}
\end{eqnarray}
~~Now by using the position representation of GUP framework introduced in (\ref{coordinatespace}), Hamiltonian changes as
\begin{eqnarray}
H(P,X)&=&\frac{P^{2}}{2m}+V(X)\nonumber\\
&=&\frac{p^{2}}{2m}+V(x)+\frac{\beta}{m}p^{4}+\mathcal{O}(\beta^{2}),
\label{H(P,X)}
\end{eqnarray}
up to first order of $\beta$. Since the GUP parameter $\beta$ is so small deformed parameter, we can use the non-degenerate perturbation theory to rewrite the above equation as
\begin{eqnarray}
H(P,X)=H(p,x)+\beta H_{p}+\mathcal{O}(\beta^{2}),
\end{eqnarray}
where $H_{p}=\dfrac{p^{4}}{m}$ is the perturbation Hamiltonian. For the energy spectrum and perturbed wave functions, one can obtain
\begin{eqnarray}
E_{n}=E_{n}^{(0)}+\beta E_{n}^{(1)}+\mathcal{O}(\beta^{2}),
\label{E}
\end{eqnarray}
and
\begin{eqnarray}
|\psi_{n}>=|\phi_{n}>+\sum_{k\neq n}C_{nk}|\phi_{k}>,
\label{state1}
\end{eqnarray}
respectively. Here,
\begin{eqnarray}
\beta E_{n}^{(1)}=<\phi_{n}|\beta H_{p}|\phi_{n}>,
\end{eqnarray}
and $|\phi_{n}>$ is the unperturbed wave function, while $C_{nk}$ is given by
\begin{eqnarray}
C_{nk}=\dfrac{<\phi_{k}|\beta H_{p}|\phi_{n}>}{E_{n}^{(0)}-E_{k}^{(0)}}.
\label{state2}
\end{eqnarray}
~~~The general Berry phase relation, driven in usual quantum mechanics with adiabatic route, is given in \cite{berryphase} by
\begin{eqnarray}
\eta_{B}=i\oint dR <\psi_{n}(R)| \nabla_R|\psi_{n}(R)>,
\label{eta}
\end{eqnarray}
where $|\psi_{n}(R)>$ and $ \nabla_R$ are the quantum state and the gradient operator with respect of the parameter space $R$, respectively. Substituting Eq.~(\ref{state1}) in the general form of Berry phase on the right-hand side of Eq.~(\ref{eta}), one can find the correction of Berry phase due to GUP as:
\begin{eqnarray}
\eta_{B}^{GUP}&=&i\oint dR <\phi_{n}(R)|\nabla_R|\phi_{n}(R)>\nonumber\\
&+&i\beta\oint dR\bigg(\sum_{l\neq n}C_{nl}<\phi_{n}(R)|\nabla_R|\phi_{l}(R)>\nonumber\\
&+& \sum_{k\neq n}C_{nk}<\phi_{k}(R)|\nabla_R|\phi_{n}(R)>\bigg)+\mathcal{O}(\beta^{2}).
\label{etaGUP}
\end{eqnarray}
~~The first term in Eq.~(\ref{etaGUP}) is Berry phase in the usual space and the other terms
give the correction to Berry phase due to GUP effect in the position representation (\ref{coordinatespace}). It should be noted that one can recover the usual form of phase factor (Eq.~(\ref{eta})) by imposing $\beta\longrightarrow 0$.
\section{Berry phase in the presence of spin-orbit interaction with a GUP}
~~~The spin-orbit interactions lead to geometrical spin phase shifts in conducting quantum rings. In fact, when the spin of electron moves inside a closed trajectory in the momentum space, the effective magnetic field produces Berry phase effects in a conducting ring \cite{PRL1993}. Here, we consider a conducting quantum ring in the presence of Rashba and Dresselhaus interactions, and study the effect of GUP on Berry phase of such quantum systems.
\subsection{Rashba interaction and GUP effect}
\begin{figure}[H]
	\centering
	\includegraphics[scale=0.7]{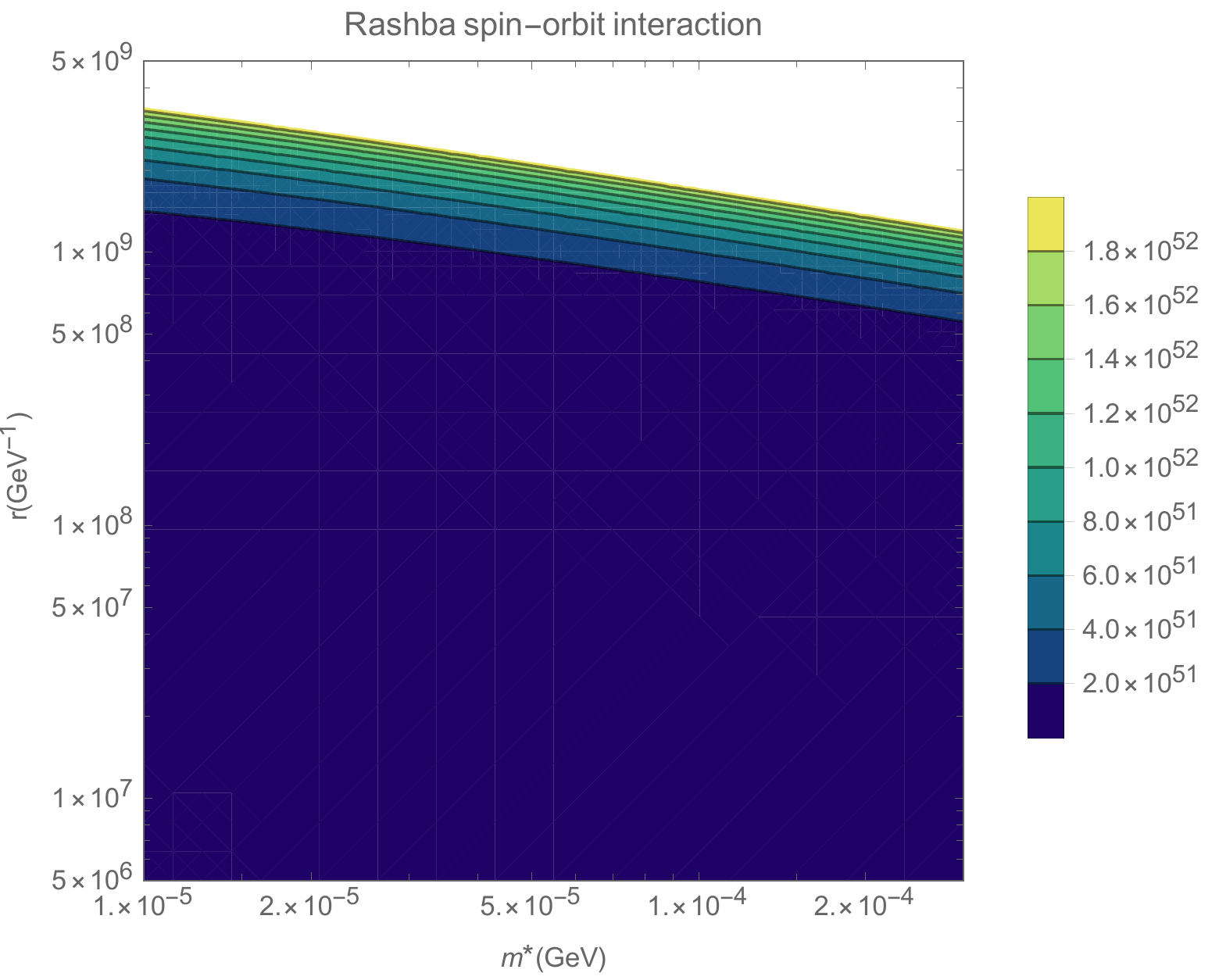}
	\caption{The upper bounds on the dimensionless GUP parameter from an electron confined in a quantum ring with space parameters ($m^{*},r$) in the presence of Rashba interaction.}
	\label{fig:Rashba}
\end{figure}

~~~The full Hamiltonian of an electron confined in a quantum ring in the presence of an external magnetic field and Rashba interaction is described in Ref. \cite{RashbaH} by
\begin{eqnarray}
H_{R}=\dfrac{(p_{x}-\frac{e}{c}A_{x})^{2}+(p_{y}-\frac{e}{c}A_{y})^{2}}{2m^{*}}+\frac{\hbar}{2}\alpha_{R}[\sigma_{x}(p_{y}-\frac{e}{c}A_{y})-\sigma_{y}(p_{x}-\frac{e}{c}A_{x})]+\hbar\omega_{B}\sigma_{z},
\label{RahbaH}
\end{eqnarray}
where $\textbf{A}$ is the vector potential and $\omega_{B}=\dfrac{egB}{2m^{*}c}$ denotes the corresponding Larmor frequency. $\alpha_{R}$ and $m^{*}$ are the strength of Rashba coupling and effective mass of electron in materials, respectively. With the polar coordinate system and the tangential component of the vector potential $A_{\varphi}=\dfrac{\Phi}{2\pi r}$, in which $\Phi$ and $r$ are the magnetic field flux and radius of ring, respectively, we get the following Hamiltonian for the electron in a closed ring
\begin{eqnarray}
H_{R}=\hbar\omega(-i\frac{\partial}{\partial\varphi}-\frac{\Phi}{\Phi_{0}})^{2}+\hbar\omega_{B}\sigma_{z}+\hbar\omega_{R}(\sigma_{x}\sin\varphi-\sigma_{y}\cos\varphi)(-i\frac{\partial}{\partial\varphi}-\frac{\Phi}{\Phi_{0}}).
\label{Rahbacly}
\end{eqnarray}
~~~Here, we defined $\omega=\dfrac{\hbar}{2m^{*}r^{2}}$ and $\omega_{R}=\dfrac{\hbar\alpha_{R}}{2r}$ as the characteristic frequencies of kinetic and Rashba terms, respectively, and $\Phi_{0}=\dfrac{2\pi\hbar c}{e}$ denotes the magnetic flux quantum. The eigenenergies of such Hamiltonian for the spin up and down are mentioned in appendix A. Relations show that, in opposite directions, we can obtain different phases due to the electron spin, a result similar to the usual AB effect ($-2\pi\dfrac{\Phi}{\Phi_{0}}$) \cite{Rashbaberry}. By using the eigenspinors of such system (appendix A), a geometrical Berry phase for an electron with state $(n=0, \lambda=+1, s=+1)$ would be obtained as
\begin{eqnarray}
\eta^{R}_{B}(\theta)&=&\pi(1+cos\theta),
\label{berry0Rashba}
\end{eqnarray}
where $\theta $ is exactly half of the solid angle subtended by the effective magnetic field.
~~~To explore the GUP effect, one can consider the position representation in the GUP modification, expressed in Eq.~(\ref{coordinatespace}), and rewrite Hamiltonian~(\ref{RahbaH}) in the form of
\begin{eqnarray}
H_{R}^{GUP}&=&H_{R}+\frac{\beta}{m^{*}}p_{x}\overrightarrow{p}^{2}(p_{x}-\frac{e}{c}A_{x})+\frac{\beta}{m^{*}}p_{y}\overrightarrow{p}^{2}(p_{y}-\frac{e}{c}A_{y})\nonumber\\
&+&\frac{\hbar}{2}\beta\alpha_{R}\overrightarrow{p}^{2}[p_{y}\sigma_{x}-p_{x}\sigma_{y}]+\mathcal{O}(\beta^{2}),
\label{HRGUP}
\end{eqnarray}
in which $\overrightarrow{p}^{2}=p_{x}^{2}+p_{y}^{2}$.
The above Hamiltonian can be written as $H_{R}^{GUP}=H_{R}+\beta H_p^{R}$ where $H_{R}$ and $H_{p}^{R}$ related to unperturbed conducting ring Hamiltonian and the perturbation Hamiltonian with GUP, respectively. By using the polar coordinates, see appendix B, we rewrite $H_{p}^{R}$ as
\begin{eqnarray}
H_{p}^{R}=\dfrac{\hbar^{4}}{m^{*}r^{4}}\{(\dfrac{\partial}{\partial \varphi})^{4}-i\dfrac{\Phi}{\Phi_{0}}(\dfrac{\partial}{\partial \varphi})^{3}\}+\dfrac{i\hbar^{4}\alpha_{R}}{2r^{3}}(\dfrac{\partial}{\partial \varphi})^{3}(\sigma_{x}\sin\varphi-\sigma_{y}\cos\varphi).
\label{HPRahbacly}
\end{eqnarray}
~~~As it is shown in appendix B, the geometrical phase under the shadow of GUP at lower perturbation order is found in form as:
\begin{eqnarray}
\eta_{B}^{R,GUP}(\theta)\approx\eta_{B}^{R}(\theta)+\frac{2\beta\pi\hbar^{2}}{r^{2}}\frac{[1+\cos\theta][3+\cos\theta]}{2+\sqrt{1+\tan^{2}\theta}},
\label{BerryRahbaGUP}
\end{eqnarray}
where the angle $\theta$ given by $\tan\theta=\alpha_{R}m^{*}r$. Here, the first term denotes the Berry phase in usual space (\ref{berry0Rashba}), the second term refers to the correction Berry phase in the presence of GUP, and thus, by considering the $\beta\rightarrow 0$ limit, we can get Eq.~(\ref{berry0Rashba}) as a desired property. Since the accuracy of current apparatus to measure a geometrical phase is less than $10^{-4}~ \text{rad}$ \cite{expberryphase}, a reasonable expectation is 
\begin{eqnarray}
|\Delta \eta_{B}^{R,GUP}(\theta)|<10^{-4}~\text{rad}~,
\end{eqnarray}
leading to the upper bound on GUP parameter in the natural unit ($\hbar=c=1$) as
\begin{eqnarray}
\beta<\dfrac{r^{2}(2+\sqrt{1+\tan^{2}\theta})}{2\pi[1+\cos\theta][3+\cos\theta]}10^{-4}.
\end{eqnarray}	
One can obtain the bound on GUP and the dimensionless parameters as $\beta<10^{8}~\rm GeV^{-2}$ and $\beta_{0}< 10^{46}$, respectively, by considering $m^{*}\simeq 0.02~m_{e}$, $r\simeq nm$, and $\alpha_{R}\simeq 10^{-4}$, where $\tan\theta=\alpha_{R}m^{*}r\simeq 5\times10^{-3}$. In Fig. (\ref{fig:Rashba}), for different materials with radius in the range of micrometer to nanometer ($\rm \mu m$ to nm), and the electron effective mass in the range of 0.02$~m_{e}$ to 0.6$~m_{e}$, the dimensionless GUP parameter ($\beta_{0}$) is displayed. One can see that the allowed region for upper bound on GUP parameter would be obtained for a ring size in the  order of nanometer where $m^{*}$ in the range of (0.02 - 0.6)$~m_{e}$.
\subsection{Dresselhaus interaction and GUP effect}
\begin{figure}[H]
	\centering
	\includegraphics[scale=0.7]{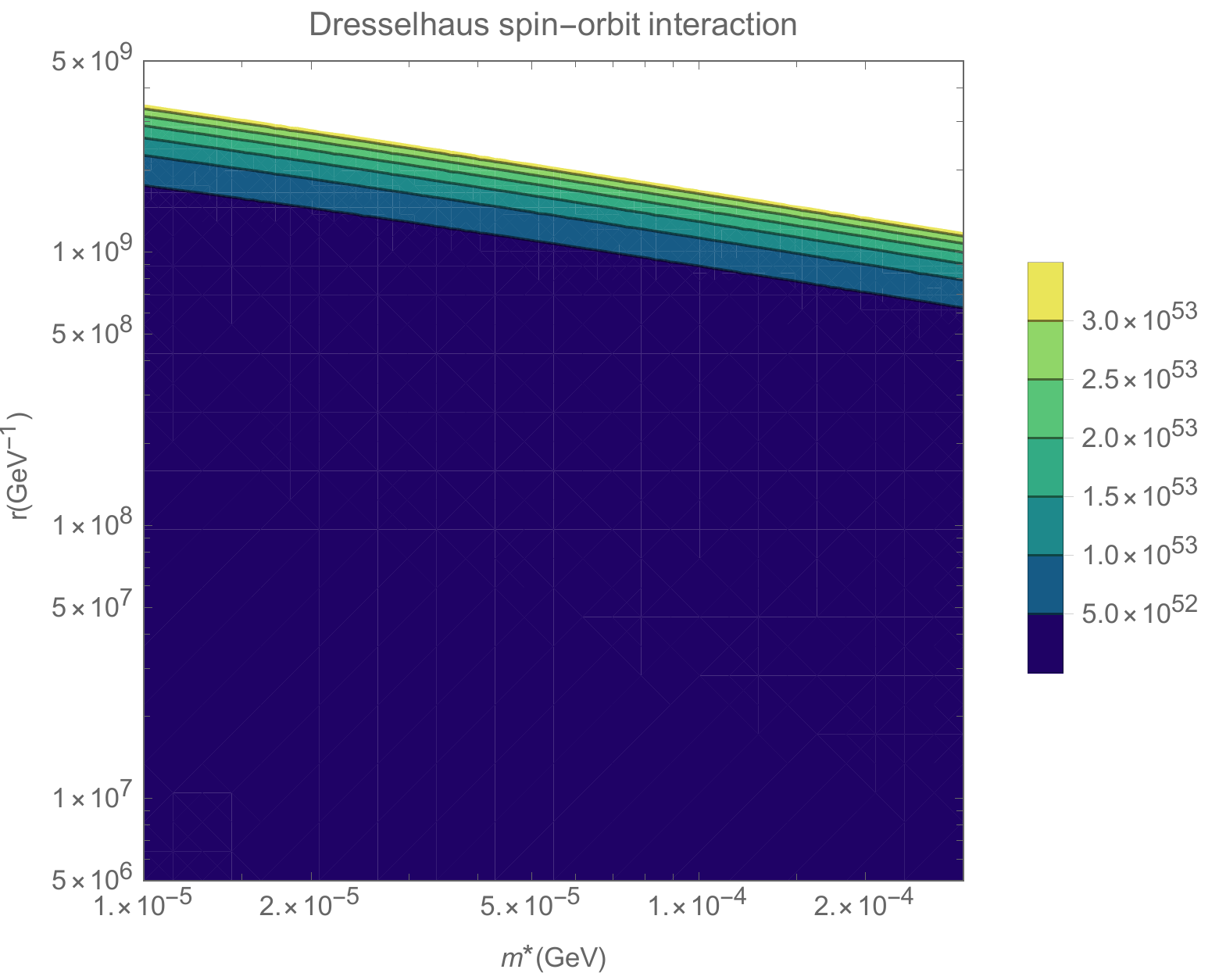}
	\caption{The upper bounds on the dimensionless GUP parameter from an electron confined in a quantum ring with space parameters ($m^{*},r$) in the presence of Dresselhaus interaction.}
	\label{fig:Dress}
\end{figure}

~~~It is possible to show the similar effect for the Dresselhaus interaction by considering a two-dimensional electron gas (2DEG) in a semiconductor. We consider a semiconductor with a normal $A_3B_5$ crystal where z-axis is parallel to the interface of a plane with Miller index $(001)$ for a rectangular quantum well and an external magnetic field (B), while the corresponding Hamiltonian \cite{PRL1993} takes the form
\begin{eqnarray}
H_{D}=\dfrac{(p_{x}-\frac{e}{c}A_{x})^{2}+(p_{y}-\frac{e}{c}A_{y})^{2}}{2m^{*}}+\frac{\hbar}{2}\alpha_{D}[\sigma_{x}(p_{x}-\frac{e}{c}A_{x})-\sigma_{y}(p_{y}-\frac{e}{c}A_{y})]+\hbar\omega_{B}\sigma_{z},
\label{HDress}
\end{eqnarray}
where $\alpha_{D}$ denotes the strength of Dresselhaus coupling. In the polar coordinates, one can gets
\begin{eqnarray}
H_{D}=\hbar\omega(-i\frac{\partial}{\partial\varphi}-\frac{\Phi}{\Phi_{0}})^{2}+\hbar\omega_{B}\sigma_{z}+\hbar\omega_{D}(\sigma_{x}\cos\varphi-\sigma_{y}\sin\varphi)(-i\frac{\partial}{\partial\varphi}-\frac{\Phi}{\Phi_{0}}),
\label{HDressrectal}
\end{eqnarray}
in which $\omega=\dfrac{\hbar}{2m^{*}r^{2}}$ and $\omega_{D}=\dfrac{\hbar\alpha_{D}}{2r}$. For $\lambda=+1$, $s=+1$, the usual Berry phase with the eigenspinors, introduced in appendix A, is given by
\begin{eqnarray}
\eta^{D}_{B}(\theta)=-\pi[2n-1+\cos\theta].
\label{Berry0Dress}
\end{eqnarray}
For the GUP Hamiltonian in the position representation, one reaches at
\begin{eqnarray}
H_{D}^{GUP}&=&H_{D}+\frac{\beta}{m^{*}}p_{x}\overrightarrow{p}^{2}(p_{x}-\frac{e}{c}A_{x})+\frac{\beta}{m^{*}}p_{y}\overrightarrow{p}^{2}(p_{y}-\frac{e}{c}A_{y})\nonumber\\
&+&\frac{\hbar}{2}\beta\alpha_{D}\overrightarrow{p}^{2}[p_{x}\sigma_{x}-p_{y}\sigma_{y}]+\mathcal{O}(\beta^{2}).
\label{HDGUP}
\end{eqnarray}
We can also write $H_{D}^{GUP}=H_{D}+\beta H_p^{D}$ in which $H_{D}$ and $H_{p}^{D}$ are related to the ring Hamiltonian with Dresselhaus coupling (Eq.~\ref{HDress}) and the perturbation Hamiltonian under the effect of GUP, respectively. The latter can finally be written as
\begin{eqnarray}
H_{p}^{D}=\dfrac{\hbar^{4}}{m^{*}r^{4}}\{(\dfrac{\partial}{\partial \varphi})^{4}-i\dfrac{\Phi}{\Phi_{0}}(\dfrac{\partial}{\partial \varphi})^{3}\}+\dfrac{i\hbar^{4}\alpha_{D}}{2r^{3}}(\dfrac{\partial}{\partial \varphi})^{3}(\sigma_{x}\cos\varphi-\sigma_{y}\sin\varphi).
\label{HPDahbacly}
\end{eqnarray}
By doing some straightforward calculations, see appendix B, we get
\begin{eqnarray}
\eta_{B}^{D,GUP}(\theta)&\approx&\eta_{B}^{D}(\theta)+\frac{4\beta\pi\hbar^{2}}{r^{2}}\frac{[1-\cos\theta][1+\cos\theta]}{3+\sqrt{9+8\tan^{2}\theta}},
\label{BerryDreslGUP}
\end{eqnarray}
as the GUP corrected Berry phase in the presence of Dresselhaus interaction. It is seen the usual Berry phase is obtained by setting $\beta\longrightarrow 0$, and the GUP effect is obtained with $\alpha_{D}\longrightarrow 0$. With respect to the condition
\begin{eqnarray}
|\Delta \eta_{B}^{D,GUP}|<10^{-4}~\text{rad},
\end{eqnarray}
where we find that the GUP parameter should satisfy the below condition
\begin{eqnarray}
\beta<\dfrac{r^{2}(3+\sqrt{9+8\tan^{2}\theta})}{4\pi[1-\cos\theta][1+\cos\theta]}10^{-4}.
\end{eqnarray}
~~Here, $\tan\theta=\alpha_{D}m^{*}r\simeq 5\times10^{-3}$ denotes a small angle which is in order of $\theta$=0.005. Therefore, bounds on GUP and dimensionless parameters are obtained as $\beta<10^{13}~\rm GeV^{-2}$ and $\beta_{0}< 10^{51}$, respectively, with $m^{*}\simeq 0.02~m_{e}$, $r\simeq nm$ and $\alpha_{D}\simeq 10^{-4}$. Fig.~(\ref{fig:Dress}) illustrates the best upper bound on GUP parameter for different materials in the Logarithmic space ($m^{*}, r$), same range as Rashba interaction where the Dresselhaus coupling is large. In the present of Dresselhaus effect, by using the current accuracy of the Berry phase detectors, the allowed region for the upper bound of GUP parameter has been depicted in Fig.~(\ref{fig:Dress}) for different materials.
\section{Conclusion}
~~~In this paper, employing position representation of GUP and perturbation theory, we could find a modification to Berry phase due to the presence of GUP and up to the first order of perturbation. Thereafter, focusing on Rashba and Dresselhaus interactions of an electron in a quantum ring, the effects of GUP on corresponding Hamiltonian and consequently, the modifications to Berry phase of each cases were obtained. With the help of the current accuracy of measurement apparatus in detecting Berry phase, we could also obtain a common upper bounds as $\beta<10^{8} ~\rm GeV^{-2}$ and $\beta<10^{13}$ from Rashba and Dresselhaus interactions, respectively. Although these bounds is in agreement with some previous works such as  electroweak measurements \cite{Das}, it is still far from the minimum upper bound $10^6$ addressed in \cite{Bushev2019}. Anyway, it will be improved if the accuracy of measurements is increased and upgraded. 
\begin{appendix}
\section{Eigenenergies, Eigenspinors and Spin-orbit interaction}
~~~In the current appendix, we address the eigenenergies and eigenspinors of an electron confined in a quantum ring with spin-orbit interactions.
\begin{itemize}
	\item Rashba interaction
\end{itemize}
~~~The 2D Hamiltonian for an electron with the effective mass $m^{*}$ subjected to Rashba coupling with constant $\alpha_{R}$ is introduced in Eq.~(\ref{RahbaH}). This Hamiltonian is not diagonal in the spin space. After diagonalization, the eigenvalues and the corresponding eigenfunctions would be listed as \cite{Rashbaberry}
\begin{eqnarray}
E_{ns}^{(R,0)}=\hbar\omega(l^{2}+\frac{1}{4})+s\hbar\sqrt{\omega_{R}^{2}l^{2}+(\omega_{B}-l\omega)^{2}},
\label{Rashbaenergy}
\end{eqnarray}
where $l=\lambda n+\frac{1}{2}+\frac{\Phi}{\Phi_{0}}$, $n$ is an integer number, $\lambda$ and $s$ denotes the travel direction and spin numbers for orbital quantum number $n\geq0$, respectively. The normalized eigenspinors, corresponding to the eigenenergies, are
\begin{eqnarray}
\phi_{n,\lambda,s}(\varphi)=e^{i\lambda n \varphi}\chi_{n,\lambda,s}; ~~~  \chi_{n,\lambda,s}=\begin{bmatrix}
\chi_{1}\\ \chi_{2}e^{i\varphi}
\end{bmatrix},
\label{rashbavectors}
\end{eqnarray}
which $\chi_{n, \lambda, s}$ is the spin components. The wave functions for travel direction number $\lambda=\pm 1$ and spin quantum number $s=\pm1$, associated with spin up ($\uparrow$) and down ($\downarrow$) states are governed by
\begin{eqnarray}
\phi_{n,+,\uparrow}&=&e^{in\varphi}
\begin{bmatrix}
\sin\frac{\theta}{2}\\ \cos\frac{\theta}{2}e^{i\varphi}
\end{bmatrix},
\nonumber\\
\phi_{n,+,\downarrow}&=&e^{in\varphi}
\begin{bmatrix}
\cos\frac{\theta}{2}\\ -\sin\frac{\theta}{2}e^{i\varphi}
\end{bmatrix},
\nonumber\\
\phi_{n,-,\uparrow}&=&e^{-in\varphi}
\begin{bmatrix}
\cos\frac{\theta}{2}\\ -\sin\frac{\theta}{2}e^{i\varphi}
\end{bmatrix},
\nonumber\\
\phi_{n,-,\downarrow}&=&e^{-in\varphi}
\begin{bmatrix}
\sin\frac{\theta}{2}\\ \cos\frac{\theta}{2}e^{i\varphi}
\end{bmatrix}.
\label{waverashba}
\end{eqnarray}
~~~The general Berry phase with Rashba coupling, for both directions with applying $\tan\theta=\dfrac{\omega_{R}}{\omega}$, is obtained by using Eq.~(\ref{eta}) as
\begin{eqnarray}
\eta_{B}^{R}(\theta)&=&-\pi[2n-1-s\cos\theta],~~~ \lambda>0,\nonumber\\
&=&-\pi[2n+1-s\cos\theta],~~~\lambda<0.
\label{RashbaB}
\end{eqnarray}
~~~The expressions~(\ref{RashbaB}) are valid if the spin-orbit energy is larger than the Zeeman energy. We should mention that the AB phase ($\dfrac{\Phi}{\Phi_{0}}$) is ignored in this equation.
\begin{itemize}
	\item Dresselhaus interaction
\end{itemize}
~~~The 2D Hamiltonian for an electron with the effective mass $m^{*}$ subjected to Dresselhaus interaction with constant $\alpha_{D}$ is introduced in Eq.~(\ref{HDress}). By diagonalizing the matrix representation of this Hamiltonian, similar to the Rashba interaction, one can find the following energy eigenvalues
\begin{eqnarray}
E_{n,s}^{(D,0)}=\hbar\omega(l^{2}+\frac{1}{4})+s\hbar\sqrt{\omega_{D}^{2}(l^{2}-\frac{1}{4})+(\omega_{B}+l\omega)^{2}},
\label{EDress}
\end{eqnarray}
where $l=\lambda n+\frac{1}{2}+\frac{\Phi}{\Phi_{0}}$, $n$ is an integer number. The corresponding wave functions, in the presence of Dresselhaus interaction, are summarized as \cite{PRL1993}
\begin{eqnarray}
\phi_{n,+,\uparrow}&=&e^{in\varphi}
\begin{bmatrix}
-\cos\frac{\theta}{2}\\ \sin\frac{\theta}{2}e^{-i\varphi}
\end{bmatrix},
\nonumber\\
\phi_{n,+,\downarrow}&=&e^{in\varphi}
\begin{bmatrix}
\sin\frac{\theta}{2}\\ \cos\frac{\theta}{2}e^{-i\varphi}
\end{bmatrix},
\nonumber\\
\phi_{n,-,\uparrow}&=&e^{-in\varphi}
\begin{bmatrix}
\sin\frac{\theta}{2}\\ \cos\frac{\theta}{2}e^{-i\varphi}
\end{bmatrix},
\nonumber\\
\phi_{n,-,\downarrow}&=&e^{-in\varphi}
\begin{bmatrix}
-\cos\frac{\theta}{2}\\ \sin\frac{\theta}{2}e^{-i\varphi}
\end{bmatrix}.
\label{Dressvectors}
\end{eqnarray}
~~~Hence, for an electron with spin up (down) in a ring, the geometrical Berry phase would be evaluated as
\begin{eqnarray}
\eta_{B}^{D}(\theta)&=&-\pi[2n-1+s\cos\theta],~~~ \lambda>0,\nonumber\\
&=&+\pi[2n+1+s\cos\theta],~~~ \lambda<0.
\end{eqnarray}
\section{Modified Berry phase with GUP}
~~~In this section, we derive the geometrical phase in the presence of GUP for an electron that is confined inside a quantum ring with spin-orbit interaction.
\begin{itemize}
	\item Rashba interaction
\end{itemize}
~~~Eq.~(\ref{HRGUP}) shows the corresponding Hamiltonian of an electron in the presence of Rashba interaction and GUP ($H_{R}^{GUP}$), where the coordinate representation in the GUP formation is applied. For investigating quantum effects of this system, it is convenient to use the polar coordinates ($r$, $\varphi$) in the following forms
\begin{eqnarray}
(p-\frac{e}{c}A)_{x}=\frac{\hbar}{r}\cos\varphi(-i\frac{\partial}{\partial \varphi}-\frac{\Phi}{\Phi_{0}}),\nonumber\\
(p-\frac{e}{c}A)_{y}=\frac{\hbar}{r}\sin\varphi(-i\frac{\partial}{\partial \varphi}-\frac{\Phi}{\Phi_{0}}).
\label{polarcor}
\end{eqnarray}
~~~Therefore, by substituting~(\ref{polarcor}) in~(\ref{HRGUP}), the perturbation term in $H_{R}^{GUP}=H_{R}+H_{p}^{GUP}$ becomes
\begin{eqnarray}
H_{p}^{R}=\dfrac{\hbar^{4}}{m^{*}r^{4}}\{(\dfrac{\partial}{\partial \varphi})^{4}-i\dfrac{\Phi}{\Phi_{0}}(\dfrac{\partial}{\partial \varphi})^{3}\}+\dfrac{i\hbar^{4}\alpha_{R}}{2r^{3}}(\dfrac{\partial}{\partial \varphi})^{3}(\sigma_{x}\sin\varphi-\sigma_{y}\cos\varphi).
\label{HpR}
\end{eqnarray}
~~~To find the modified Berry phase in the presence of GUP, we expand the expression~(\ref{etaGUP}) in the parameter space $\varphi$ as follows
\begin{eqnarray}
\eta_{B}^{R,GUP}&=& \eta_{B}^{R}(\theta)
+2i\int_{0}^{2\pi} d\varphi \sum_{k\neq n}C^{R}_{nk}<\phi_{n}(\varphi)|\frac{\partial}{\partial \varphi}|\phi_{k}(\varphi)>+\mathcal{O}(\beta^{2}),
\label{eq1}
\end{eqnarray}
where 
\begin{eqnarray}
C^{R}_{nk}=\dfrac{<\phi_{k}(\varphi)| \beta H^{R}_{p}|\phi_{n}(\varphi)>}{E_{n}^{(R,0)}-E_{k}^{(R,0)}}.
\label{CnkR}
\end{eqnarray}
~~~It should be noted that the states of system,   $\phi_{n}(\varphi)\equiv \phi_{n,\lambda,s}(\varphi)$ are introduced in appendix A for Rashba interaction. For simplicity, we suppose  that the initial state of system would be as $(n,\lambda=+1, s=+1)$, then Eq.~(\ref{eq1}) can be written as
\begin{eqnarray}
\eta_{B}^{R,GUP}&=& \eta_{B}^{R}(\theta)
+2i\beta\int_{0}^{2\pi} d\varphi \sum_{k\neq n}\dfrac{<\phi_{k,\lambda',s'}(\varphi)|H_{p}^{R}|\phi_{n,+1,+1}(\varphi)>}{E_{n,+1,+1}^{(R,0)}-E_{k,\lambda',s'}^{(R,0)}}\times\nonumber\\
&&<\phi_{n,+1,+1}(\varphi)|\frac{\partial}{\partial \varphi}|\phi_{k,\lambda',s'}(\varphi)>,
\label{eq2}
\end{eqnarray}
where by considering all intermediate states $(\lambda', s')$, the non-zero terms can be found in the form of
\begin{eqnarray}
\eta_{B}^{R,GUP}&=& \eta_{B}^{R}(\theta)
+2i\beta\int_{0}^{2\pi} d\varphi \sum_{k\neq n}\dfrac{<\phi_{k,+1,+1}(\varphi)|\frac{\hbar^{4}}{m^{*}r^{4}}(\frac{\partial}{\partial \varphi})^{4}|\phi_{n,+1,+1}(\varphi)>}{E_{n,+1,+1}^{(R,0)}-E_{k,+1,+1}^{(R,0)}}\times\nonumber\\
&&<\phi_{n,+1,+1}(\varphi)|\frac{\partial}{\partial \varphi}|\phi_{k,+1,+1}(\varphi)>.
\label{eq3}
\end{eqnarray}
~~~It should also be noted that the second term in $H_{p}^{R}$ have no contribution due to the orthogonal condition in the eigenspinors. Here, the assumption $\frac{\Phi}{\Phi_{0}}\longrightarrow 0$ is applied and consequently, the numerator and denominator of Eq.~(\ref{eq3}) can be simplified by using the eigenspinors~(\ref{rashbavectors}) and eigenenergies~(\ref{Rashbaenergy}), in the Rashba part of appendix A, respectively, as
\begin{eqnarray}
&&e^{-ik\varphi}
\begin{bmatrix}
\sin\frac{\theta}{2}& \cos\frac{\theta}{2}e^{-i\varphi}
\end{bmatrix}(\frac{\partial}{\partial \varphi})^{4}
\begin{bmatrix}
\sin\frac{\theta}{2}\\ \cos\frac{\theta}{2}e^{i\varphi}
\end{bmatrix}e^{in\varphi}\times\nonumber\\
&&e^{-in\varphi}
\begin{bmatrix}
\sin\frac{\theta}{2}& \cos\frac{\theta}{2}e^{-i\varphi}
\end{bmatrix}(\frac{\partial}{\partial \varphi})
\begin{bmatrix}
\sin\frac{\theta}{2}\\ \cos\frac{\theta}{2}e^{i\varphi}
\end{bmatrix}e^{ik\varphi}
\nonumber\\
&&=i\{n^{4}+(6n^{2}+4n^{3}+4n+1)\cos^{2}\frac{\theta}{2}\}\{\frac{2k+1+\cos\theta}{2}\},
\label{num}
\end{eqnarray} 
and
\begin{eqnarray}
E_{n,+1,+1}^{(R,0)}-E_{k,+1,+1}^{(R,0)}= \hbar\omega\{l^{2}-l'^{2}+(l-l')\sqrt{1+\tan^{2}\theta}\},
\label{deno}
\end{eqnarray}
where we set $\omega_{B}=0~(\Phi\longrightarrow 0)$, $\tan\theta=\frac{\omega_{R}}{\omega}=\alpha_{R}m^{*}r$, $l=n+\frac{1}{2}$, and $l'=k+\frac{1}{2}$. By doing some straightforward calculations and taking integral on $\varphi$ space, Eq.~(\ref{eq3}) leads to the following expansion
\begin{eqnarray}
\eta_{B}^{R,GUP}(\theta)&=&\eta_{B}^{R}(\theta)+\frac{2\beta \pi \hbar^{2}}{r^{2}}[\frac{(1+\cos\theta)(3+\cos\theta)}{2+\sqrt{1+\tan^{2}\theta}}+\frac{(1+\cos\theta)(5+\cos\theta)}{6+2\sqrt{1+\tan^{2}\theta}}\nonumber\\
&+&\frac{(1+\cos\theta)(7+\cos\theta)}{12+3\sqrt{1+\tan^{2}\theta}}+...],
\label{etaBRashba}
\end{eqnarray}
in which we set $n=0$, $k=1,2,...$, and moreover, the $\alpha_R\rightarrow 0$ limit addresses the effects of GUP on Berry phase in the absence of Rashba interaction. One can observe in Eq.~(\ref{etaBRashba}), the terms with both Rashba and GUP corrections are decreasing in a given angle while the first term has a larger amplitude in comparison to others. The signification effects in Rashba interaction and GUP effects at lower states may take place in the nearest states. Hence, we consider only the first term ($k=1$) to generate an effective contribution for examining the GUP parameter, where the corresponding Berry phase can be written as
\begin{eqnarray}
\eta_{B}^{R,GUP}(\theta)\approx\eta_{B}^{R}(\theta)+\frac{2\beta\pi\hbar^{2}}{r^{2}}\frac{[1+\cos\theta][3+\cos\theta]}{2+\sqrt{1+\tan^{2}\theta}}.
\label{etaBRashba2}
\end{eqnarray}
The first term refers to Berry phase in the presence of Rashba coupling as
\begin{eqnarray}
\eta_{B}^{R}(\theta)=i\int_{0}^{2\pi} d\varphi <\phi_{0,+1,+1}(\varphi)|\frac{\partial}{\partial \varphi}|\phi_{0,+1,+1}(\varphi)>=\pi[1+\cos\theta],
\label{44}
\end{eqnarray}
which is introduced in appendix A with setting $n=0$, $\lambda=+1$, and $s=+1$. It should be noted that the modification of Berry phase in Eq.~(\ref{etaBRashba}) shows that the different order of magnitude in GUP parameter ($\beta$) may be testable by the variety of experiments.\\
\begin{itemize}
	\item Dresselhaus interaction
\end{itemize}
~~~To find the modified Berry phase in the presence of Dresselhaus interaction and GUP, one can follow similar way as Rashba interaction with using the corresponding expression in Dresselhaus interaction. We start with the corresponding Hamiltonian which is introduced in Eq.~(\ref{HDGUP}). We can rewrite this Hamiltonian in the polar coordinates which is addressed in Eq.~(\ref{polarcor}) as follows
\begin{eqnarray}
H_{p}^{D}=\dfrac{\hbar^{4}}{m^{*}r^{4}}\{(\dfrac{\partial}{\partial \varphi})^{4}-i\dfrac{\Phi}{\Phi_{0}}(\dfrac{\partial}{\partial \varphi})^{3}\}+\dfrac{i\hbar^{4}\alpha_{D}}{2r^{3}}(\dfrac{\partial}{\partial \varphi})^{3}(\sigma_{x}\cos\varphi-\sigma_{y}\sin\varphi).
\label{HpD}
\end{eqnarray}
~~~The Berry phase with GUP correction is written in the following form
\begin{eqnarray}
\eta_{B}^{D,GUP}&=& \eta_{B}^{D}(\theta)
+2i\int_{0}^{2\pi} d\varphi \sum_{k\neq n}C^{D}_{nk}<\phi_{n}(\varphi)|\frac{\partial}{\partial \varphi}|\phi_{k}(\varphi)>+\mathcal{O}(\beta^{2}),
\label{eq4}
\end{eqnarray}
where 
\begin{eqnarray}
C^{D}_{nk}=\dfrac{<\phi_{k}(\varphi)| \beta H^{D}_{p}|\phi_{n}(\varphi)>}{E_{n}^{(D,0)}-E_{k}^{(D,0)}}.
\label{CnkD}
\end{eqnarray}
~~~By considering the similar conditions for quantum system in the presence of Dresselhaus interaction $(n,\lambda=+1, s=+1)$, the non-zero terms are given by
\begin{eqnarray}
\eta_{B}^{D,GUP}&=& \eta_{B}^{D}(\theta)
+2i\beta\int_{0}^{2\pi} d\varphi \sum_{k\neq n}\dfrac{<\phi_{k,+1,+1}(\varphi)|\frac{\hbar^{4}}{m^{*}r^{4}}(\frac{\partial}{\partial \varphi})^{4}|\phi_{n,+1,+1}(\varphi)>}{E_{n,+1,+1}^{(D,0)}-E_{k,+1,+1}^{(D,0)}}\times\nonumber\\
&&<\phi_{n,+1,+1}(\varphi)|\frac{\partial}{\partial \varphi}|\phi_{k,+1,+1}(\varphi)>.
\label{eq5}
\end{eqnarray}
~~~The numerator and denominator of Eq.~(\ref{eq5}) can be found by using the eigenspinors and eigenenergies which are mentioned in the part of Dresselhaus of appendix A, respectively, as
\begin{eqnarray}
&&e^{-ik\varphi}
\begin{bmatrix}
-\cos\frac{\theta}{2}& \sin\frac{\theta}{2}e^{i\varphi}
\end{bmatrix}(\frac{\partial}{\partial \varphi})^{4}
\begin{bmatrix}
-\cos\frac{\theta}{2}\\ \sin\frac{\theta}{2}e^{-i\varphi}
\end{bmatrix}e^{in\varphi}\times\nonumber\\
&&e^{-in\varphi}
\begin{bmatrix}
-\cos\frac{\theta}{2}& \sin\frac{\theta}{2}e^{i\varphi}
\end{bmatrix}(\frac{\partial}{\partial \varphi})
\begin{bmatrix}
-\cos\frac{\theta}{2}\\ \sin\frac{\theta}{2}e^{-i\varphi}
\end{bmatrix}e^{ik\varphi}
\nonumber\\
&&=i\{n^{4}+(6n^{2}-4n^{3}-4n+1)\sin^{2}\frac{\theta}{2}\}\{\frac{2k-1+\cos\theta}{2}\},
\label{numD}
\end{eqnarray} 
and
\begin{eqnarray}
E_{n,+1,+1}^{(D,0)}-E_{k,+1,+1}^{(D,0)}&=&
\hbar\omega\{l^{2}-l'^{2}+\sqrt{\tan^{2}\theta(l^{2}-\frac{1}{4})+l^{2}}\nonumber\\
&-&\sqrt{\tan^{2}\theta(l'^{2}-\frac{1}{4})+l'^{2}}\}.
\label{denoD}
\end{eqnarray}
~~Here, we apply $\tan\theta=\frac{\omega_{D}}{\omega}$.
The GUP correction on Berry phase in the presence of Dresselhaus interaction can be driven easily by doing some straightforward calculations as
\begin{eqnarray}
\eta_{B}^{D,GUP}(\theta)&=&\eta_{B}^{D}(\theta)+\frac{4\beta\pi\hbar^{2}}{r^{2}}[\frac{(1-\cos\theta)(1+\cos\theta)}{3+\sqrt{9+8\tan^{2}\theta}}\nonumber\\
&+&\frac{(1-\cos\theta)(3+\cos\theta)}{11+\sqrt{25+24\tan^{2}\theta}}
+\frac{(1-\cos\theta)(5+\cos\theta)}{23+\sqrt{49+48\tan^{2}\theta}}+...],
\label{etaBD}
\end{eqnarray}
where we set $n=0$ and $k=1,2,3,...$. By considering the effective contribution for GUP effect, where the state $k=1$ provides the larger contribution in compare of other terms as mentioned in Rashba interaction, one can obtain 
\begin{eqnarray}
\eta_{B}^{D,GUP}(\theta)&\approx&\eta_{B}^{D}(\theta)+\frac{4\beta\pi\hbar^{2}}{r^{2}}\frac{[1-\cos\theta][1+\cos\theta]}{3+\sqrt{9+8\tan^{2}\theta}},
\label{etaD2}
\end{eqnarray}
where angle $\theta$ refers to $\tan\theta=\alpha_{D}m^{*}r$. The first term refers to Berry phase in the presence of Dresselhaus interaction as shown in appendix A, where the settings $n=0$, $\lambda=+1$, and $s=+1$ are used.
\begin{eqnarray}
\eta_{B}^{D}(\theta)=i\int_{0}^{2\pi} d\varphi <\phi_{0,+1,+1}(\varphi)|\frac{\partial}{\partial \varphi}|\phi_{0,+1,+1}(\varphi)>=\pi[1-\cos\theta].
\label{55}
\end{eqnarray}
\end{appendix}

\end{document}